\documentclass[journal=jctcce]{achemso}
\setkeys{acs}{articletitle = true}
\pdfoutput=1 
\usepackage[utf8]{inputenc}
\usepackage[T1]{fontenc}
\usepackage{amssymb}
\usepackage{amsmath}
\usepackage{simplewick}
\usepackage{graphics}
\usepackage{color}
\usepackage{calc}

\newcommand{\Tr}{\operatorname{Tr}}
\renewcommand{\d}{\textrm{d}}
\renewcommand{\vec}[1]{\mathbf{#1}}
\newcommand{\e}{{\mathrm{e}}}

\newcommand{\diagramBox}[2][0.5]{\raisebox{0.5ex-#1\height}{#2}}

\newcommand{\SE}{Schr\"odinger equation}
\newcommand{\Eq}[1]{Eq.~\eqref{#1}}
\newcommand{\Fig}[1]{FIG.~\ref{#1}}

\title{
  Finite temperature coupled cluster theories for extended systems
}

\author{Felix Hummel}
\email{felix.hummel@tuwien.ac.at}
\affiliation{
  Institute for Theoretical Physics, TU Wien,\\
  Wiedner Hauptstraße 8-10/136, 1040 Vienna, Austria
}
\date{September 27, 2018, PREPRINT}


\begin{document}

\begin{abstract}
At zero temperature, coupled cluster theory is widely used to predict
total energies, ground state expectation values and even excited states
for molecules and extended systems.
However, for systems with a small band gap, such as metals,
the zero-temperature approximation not necessarily holds.
Thermal effects may even give rise to interesting chemistry on metal surfaces.
Most approaches to temperature dependent electronic properties employ
finite temperature perturbation theory in the Matsubara frequency formulation.
Computations require a large number of Matsubara frequencies
to yield sufficiently accurate results, especially at low temperatures.

This work, and independently the work of White and Chan,\cite{white_2018}
proposes a coupled cluster
implementation directly in the imaginary time domain on the compact
interval $[0,\beta]$, closely related to
the thermal cluster cumulant approach of
Mukherjee and coworkers.\cite{sanyal_1992,sanyal_1993,mandal_2003}
Here, the arising imaginary time dependent coupled cluster amplitude integral
equations are solved in the linearized direct ring doubles approximation,
also referred to as Tamm--Dancoff approximation with second order
(linearized) screened exchange.
In this framework, the transition from finite to zero temperature is uniform
and comes at no extra costs, allowing to go to temperatures as low
as room temperature.
In this approximation, correlation grand potentials are
calculated over a wide range of temperatures for solid lithium,
a metallic system, and for solid silicon, a semiconductor.
\end{abstract}

\maketitle

\section{Background}
The many-body \SE\ for the electrons in matter can only be solved approximately
for systems having more than one electron. Many approximation methods exist,
ranging from fast methods providing low accuracy, such as orbital-free density
functional theory, to slow methods providing high accuracy, such as
full configuration interaction quantum Monte Carlo. Coupled cluster
methods provide relatively high accuracy at computational costs that scale
only polynomially with the size of the system to be
calculated.\cite{coester_short-range_1960,cizek_use_1969}
Although the scaling is still steep, modern computer
facilities enable coupled cluster calculations for systems large enough
to extrapolate to the infinite size of a solid or a
surface.\cite{gruber_2018}
Moreover, coupled cluster methods offer a whole hierarchy of approximations
having increasing accuracy with increasing computational costs. Within this
hierarchy, coupled cluster singles doubles with perturbative triples (CCSD(T))
is regarded accurate enough for most practical purposes, providing an
accuracy of about 1\,kcal/mol or roughly
40\,meV/atom.\cite{shavitt_many-body_2009}

Coupled cluster methods are well-proven for the zero-temperature case
where the underlying starting point calculation, a Hartree--Fock or
density functional theory calculation, is non-degenerate. For molecules
the energy spectrum is discrete and gapped, making a
zero-temperature method a valid approximation. Degeneracies can still occur and
and they may need to be treated fully quantum-mechanically, as done
by multi reference methods. There are various forms of coupled cluster
theories for this case. The methods are, however, not as settled as in the
non-degenerate closed-shell case and are thus subject to ongoing research,
but not scope of this work.\cite{coughtrie_2018}
Metallic systems, on the other hand, have a dense energy spectrum such that
they interact with their decoherent environment even at the lowest energy
scales.
They need to be treated by a theory capable of describing their
decoherent state being a mixture, rather than a superposition of pure
quantum mechanical states.

At zero temperature the electrons of matter assume the state of lowest
possible energy. At finite temperature the many-body quantum system of
electrons can be found in any of its states. At thermal equilibrium the
probability of finding it in an eigenstate with energy $E_n$ is proportional
to the Boltzmann factor $\e^{-\beta E_n}$, where $\beta=1/k_\mathrm{B}T$
denotes the inverse temperature of the system.\cite{breuer_2007}
If the system can exchange
electrons with the bath in addition to energy, it can also be found
in states with fewer or more electrons than suggested, say, by the
number of protons $N$.
Thus, a many-body electronic system at finite temperature can be computed by,
first, calculating the relevant many-body states $|\Psi_n\rangle$ with a
zero-temperature theory which is able to yield excited states,
and then determine the finite temperature many-body system by the
mixture of states $|\Psi_n\rangle$ with probabilities proportional to their
respective Boltzmann factor $\e^{-\beta E_n}$.

\subsection{Correlating large systems at finite temperature}
\begin{figure}[h]
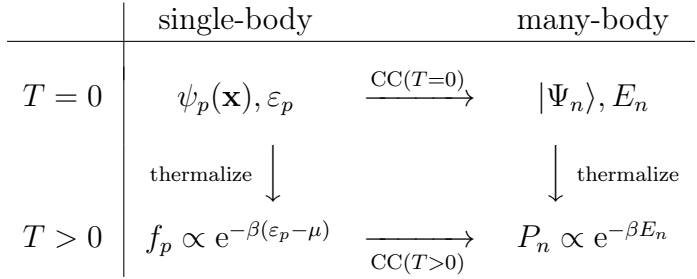

\begin{center}
\begin{tabular}{c|ccc}
  & single-body &  & many-body \\\hline \\[-1ex]
  $T=0$\, &
    $\psi_p(\vec x),\varepsilon_p$ &
    $\xrightarrow{\mathrm{CC}(T=0)}$ &
    $|\Psi_n\rangle,E_n$ \\[2ex]
  &
    {\scriptsize thermalize}
    $\Big\downarrow$
    {\scriptsize\hspace{3ex}} &
    &
    {\scriptsize\hspace{3ex}}
    $\Big\downarrow$
    {\scriptsize thermalize} \\[2ex]
  $T>0$\, &
    \,$
      f_p \propto \e^{-\beta(\varepsilon_p-\mu)}
    $ &
    $\xrightarrow[\mathrm{CC}(T>0)]{}$ &
    $P_n \propto \e^{-\beta E_n}$
\end{tabular}
\end{center}
\caption{
  Thermalizing the single-body description and adding correlation by a
  finite temperature coupled cluster formulation should agree with
  adding correlation at zero temperature and thermalizing the resulting
  many-body states in cases where both methods are applicable.
}
\label{fig:ThermalizingCorrelating}
\end{figure}

For large systems, and particularly for metals, the number of
relevant excited states quickly becomes unmanageable for practical purposes.
Furthermore, conducting the zero temperature coupled cluster calculations
is difficult at best for metallic systems due to (quasi) degeneracies
of the uncorrelated mean-field description,
such as Hartree--Fock (HF) or density functional theory (DFT),
serving as the starting point of coupled cluster theories.
Alternatively, one can already start from a thermal mean-field description,
where each single-body level $p$ forms an independent fermionic system
occupied with a probability $f_p$,
and add correlation by means of a finite temperature many-body theory.
It is the scope of this work to translate zero temperature
coupled cluster theories CC($T=0$) to a finite-temperature starting
point, circumventing the difficulties of the first approach arising in
large and metallic systems. For systems, where zero temperature
coupled cluster can be readily applied, such as insulating systems,
thermalizing and adding correlation should commute in the limit of exact
correlation theories, as laid out in \Fig{fig:ThermalizingCorrelating}.
For $T\rightarrow0$ it would even be desirable that a finite-temperature
correlation theory agrees with its zero-temperature counterpart,
irrespective of its accuracy,
in the non-degenerate case where the zero-temperature coupled cluster
theory can be applied.

\subsection{Imaginary time formulation}
\begin{figure}[h]
\begin{center}
\begin{tabular}{c|c|c|c}
    $T$ & $\tau$ integration &
  $\displaystyle\langle
      \contraction{}{\hat c}{_a}{\hat c}
        \hat c_a \hat c^\dagger_b
    \rangle_0
  $
  & diagrams \\
  &&& \\[-2.5ex]\hline
  &&& \\[-2.5ex]
  $=0$\,
    & \,$\displaystyle
      \int\limits_{
        {-\infty}<\tau_1<\ldots<\tau_{n-1}
        < {\tau_n=0}
      }
      \hspace{-3ex}
      \d\tau_1\ldots\d\tau_{n-1}
    $
    & $\delta^b_a\theta(\varepsilon_a-\mu)$
    & \diagramBox{\includegraphics{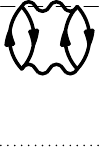}} \\
  &&& \\[-2.5ex]
  $>0$\,
    & $\displaystyle
      -\frac1{{\beta}}
      \int\limits_{
        {0}<\tau_1<\ldots<{\tau_n<\beta}}
      \hspace{-2ex}
      \d\tau_1\ldots{\d\tau_n}
    $
    & $\delta^b_a {f^a}$
    & \diagramBox{\includegraphics{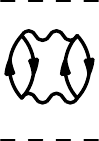}}
\end{tabular}
\end{center}
\caption{
  The many-body perturbation expansion at zero and at finite temperature
  can be given in terms of the same diagrams, differing only in the
  way contractions are evaluated and imaginary times $\tau$ are to be
  integrated.
  The dashed lines indicate $\tau=0$ and $\beta$, respectively. The
  dotted line denotes minus infinity. $f^a=(1-f_a)$.
}
\label{fig:FiniteTempMBPT}
\end{figure}

The framework employed by finite temperature many-body perturbation theory
(MBPT) provides the means of
adding correlation to a thermalized starting point.
In Appendix \ref{sec:Derivation} it is briefly outlined and
according to \Eq{eqn:FreeEnergyDifference} therein,
a given term of the perturbation expansion of $n$th order is calculated
by evaluating contractions
$
\langle
      \ldots\contraction{}{\hat c}{_a\ldots}{\hat c}
        \hat c_a \ldots\hat c^\dagger_b\ldots
    \rangle_0
$
of the perturbations $\hat H_1$ at $n$
distinct imaginary times $\tau_1,\ldots,\tau_n$,
integrating the times in an ordered fashion on the
interval $0<\tau_1<\ldots<\tau_n<\beta$. Between the times
of the perturbations, say $\tau_1$ and $\tau_2$, the system propagates
according to the mean-field description, namely
$\exp\{-(\tau_2-\tau_1)(\hat H_0 - \mu\hat N)\}$.
This is analogous to the time dependent formulation of zero temperature
many-body perturbation theory, Wick rotated to imaginary time about
the Fermi energy $\varepsilon_\mathrm{F}=\mu$.
In the zero temperature MBPT the times are integrated on the infinite interval
$-\infty<\tau_1<\ldots<\tau_n=0$. The terms of the many-body expansion are
otherwise identical and can in particular be depicted by the same diagrams.
The respective evaluation scheme is summarized in
\Fig{fig:FiniteTempMBPT}. Note, however, that there are diagrams with
vanishing contribution at zero temperature but not at finite temperature.

The zero temperature coupled cluster formalism can be stated as a
recipe to generate terms of the many-body perturbation expansion.
The expansion becomes more and more complete as one goes up in the
hierarchy of coupled cluster approximations.
Thus, zero temperature coupled cluster theories can be generalized
to finite temperature by translating their many-body expansion terms
according to \Fig{fig:FiniteTempMBPT}.

\subsection{Related work}
The generalization of Wick's theorem to mixtures, rather than pure Slater
determinants, is the key to finite temperature many-body perturbation theory.
It was brought forward by Matsubara.\cite{matsubara_new_1955,thouless_2014}
Finite temperature was introduced to mean-field theories shortly after,
first to Hartree--Fock\cite{mermin_1963} and then to density functional
theory.\cite{mermin_1965}
Being relatively fast, finite temperature Hartree--Fock and DFT
are widely used for \emph{ab-initio} calculations
and subject to ongoing research.\cite{eschrig_2010}
Second order finite temperature M\o ller--Plesset theory includes
correlation beyond DFT. It is, however, divergent in the low temperature,
large system size limit for metals.\cite{santra_2017}
Many theories exist employing Dyson-like recursion schemes to go to
infinite order in the perturbation.
The finite temperature random phase approximation (RPA) has been used
to study the warm electron
gas.\cite{gupta_1980,perrot_1982,perrot_1984,csanak_1997}
RPA corresponds to the direct term of the direct ring approximation
of coupled cluster doubles.\cite{scuseria_2008}
Self consistent Green's function methods\cite{welden_2016}
and dynamical mean field theory (DMFT)\cite{metzner_1989,held_2007}
solve a coupled set of Dyson-like of equations and have been applied to
realistic systems.

The above methods are formulated in the Matsubara frequency domain,
which requires knowledge of the symmetry factors of the included diagrams
to avoid multiple counting.\cite{mattuck_1992}
On the other hand, the recursion recipe of coupled cluster theory
generates diagrams of all kinds of symmetry, prohibiting any frequency
domain and requiring strict (imaginary) time ordering.
Hermes \emph{et al.} \cite{hermes_2015} employed a steady-state \emph{ansatz}
for coupled cluster doubles within their renormalized perturbation theory
\cite{hirata_kohnluttinger_2013}
and apply it to study the
Peierls transition in one dimensional infinite hydrogen chains.

Mukherjee and coworkers have rigorously generalized the coupled
cluster \emph{ansatz} by working within particularly contracted terms as a
chosen \emph{ansatz}, termed thermal cluster cumulant
method.\cite{sanyal_1992,sanyal_1993,mandal_2003}
Translating terms of the formal many-body perturbation expansion shows that
this \emph{ansatz} is well-justified.
The authors solve this equation for model systems, here a method is
presented to solve it for \emph{ab-initio} Hamiltonians of extended systems.
Independently of this work, White and Chan\cite{white_2018} propose
using the same imaginary time dependent formulation as proposed herein.
The authors implement full coupled cluster singles doubles and study the warm
uniform electron gas at different densities and temperatures.

Finally, fractional occupancy formulations of coupled cluster
theories\cite{yang_2013,margraf_2018}
use the same generalization of Wick's theorem to mixtures as the one
employed by finite temperature many-body perturbation theory.
However, they usually cannot be viewed as the zero-temperature limit
of the grand canonical finite-temperature formalism,
as outlined in Subsection \ref{ssc:ZeroTempLimit} and discussed in detail
in Santra and Schirmer.\cite{santra_2017}

\subsection{Structure of this work}
Section \ref{sec:ZeroTemperature} gives an imaginary time dependent
formulation of zero temperature coupled cluster theories, which also
defines a subset of the many-body perturbation expansion.
Section \ref{sec:FiniteTemperature} translates the subset using
the framework of finite temperature many-body perturbation theory.
In Section \ref{sec:Solving}
an algorithm is given to solve the occurring equations for \emph{ab-initio}
Hamiltonians.
Section \ref{sec:Results} applies finite temperature
linearized direct ring coupled cluster doubles
to solid lithium and silicon to demonstrate its applicability to metals and
its convergence behavior for low temperatures.
Appendix \ref{sec:Derivation} gives a brief derivation of finite temperature
many-body perturbation theory.

\section{
  Imaginary time dependent zero temperature coupled cluster
}
\label{sec:ZeroTemperature}
This section applies the coupled cluster \emph{ansatz} first to the
stationary, then to the (imaginary) time dependent \SE. Both formulations
yield equivalent results at zero temperature, however, the latter can
be used in a finite-temperature framework where one needs to find
the imaginary time propagator $\exp\{-\beta\hat H\}$ to the point $\beta$
in imaginary time.

\subsection{The coupled cluster \emph{ansatz}}
Coupled cluster theories start from a mean-field Hartree--Fock or DFT
calculation.
Let $\psi_p(\vec x)$ be the spin orbitals of the mean-field Hamiltonian
$\hat H_0$ and let
$|\Phi\rangle$ denote the Slater determinant of the ground state, where the
lowest $N$ orbitals are occupied by the $N$ electrons present in the system.
At zero temperature we will use the letters $i,j,k,\ldots$ to label occupied
orbitals,
$a,b,c,\ldots$ to label virtual orbitals and $p,q,r,\ldots$ to label general
spin orbitals.
Coupled cluster chooses an exponential \emph{ansatz}
acting on the mean-field Slater determinant
for the approximation of the many-body wave function
\begin{equation}
  |\Psi\rangle = \e^{\hat T} |\Phi\rangle\,.
  \label{eqn:ZeroTCC}
\end{equation}
The cluster operator $\hat T$
is expanded in excitation levels
\begin{equation}
  \hat T = \hat T_1 + \hat T_2 + \ldots
  = \sum_{ai} T^a_i \hat t^a_i
    + \sum_{abij} T^{ab}_{ij} \hat t^{ab}_{ij}
    + \ldots
\end{equation}
with the excitation operators $\hat t^a_i=\hat c^\dagger_a\hat c_i$,
$\hat t^{ab}_{ij}=\hat c^\dagger_a\hat c^\dagger_b\hat c_j\hat c_i$, \ldots, and
with the scalar arrays $T^a_i$, $T^{ab}_{ij}$, \ldots to be determined, which
are referred to as singles, doubles, \ldots amplitudes, respectively.
This choice ensures multiplicative
separability of the approximate many-body wave function for non-interacting
subsystems by construction.
In the projection coupled cluster method the amplitudes are determined by
inserting \Eq{eqn:ZeroTCC} into the stationary \SE\
$\hat H|\Psi\rangle = E|\Psi\rangle$, left-multiplying it with $\exp\{-\hat T\}$
and projecting onto excited Slater determinants $\langle\Phi|\hat t^i_a$,
$\langle\Phi|\hat t^{ij}_{ab}$, \ldots, giving
\begin{equation}
  0 = \langle \Phi | \hat t^i_a\, \e^{-\hat T} \hat H\, \e^{\hat T}
    |\Phi\rangle, \quad
  0 = \langle \Phi | \hat t^{ij}_{ab}\, \e^{-\hat T} \hat H\, \e^{\hat T}
    |\Phi\rangle,
  \quad \ldots
  \label{eqn:ZeroTCCAmplitudes}
\end{equation}
In practice, the excitation level is truncated at a given level, say
at the doubles level, and the first and the second equations are used to
solve for the singles and doubles amplitudes, respectively. The truncation
level determines the quality of the approximation.
Having found the amplitudes, the above equation is projected onto the
ground state Slater determinant $\langle\Phi|$ to yield the
coupled cluster ground state energy $E$ in the respective approximation
\begin{equation}
  E = \langle\Phi| \e^{-\hat T} \hat H\, \e^{\hat T}|\Phi\rangle.
  \label{eqn:ZeroTCCE}
\end{equation}

The expectation values of Eqs.~(\ref{eqn:ZeroTCCAmplitudes}, \ref{eqn:ZeroTCCE})
are evaluated by summing over all fully contracted terms occurring in the
expansions according to Wick's theorem.
It turns out that only terms of the positive exponential remain
where all occurring operators are connected by the contractions, denoted
by an apostrophe on the expectation value brackets $\langle\cdot\rangle'$.
The operator equation for determining the doubles amplitudes reads for instance
\begin{equation}
  0 = \langle\Phi|\hat t^{ij}_{ab} \hat H\,\e^{\hat T}|\Phi\rangle'
  \label{eqn:ZeroTConnectedDoubles}
\end{equation}
and the expansion of the remaining exponential now terminates since the number
of the contractions which can be connected to $\hat H$ and $\hat t$ is
finite for a given excitation level of the amplitudes.
The energy expression is also simplified by regarding only connected terms to
\begin{equation}
  E = \langle\Phi|\hat H\,\e^{\hat T}|\Phi\rangle'.
  \label{eqn:ZeroTConnectedEnergy}
\end{equation}

\subsection{Defining time dependent amplitudes}
In the finite temperature many-body formalism one is required
to compute a density operator $\hat\rho \propto \exp\{-\beta\hat H\}$.
This is equivalent to the time evolution operator from zero to the finite 
point $\beta$ in imaginary time.
We start by obtaining equations for the amplitudes
of the zero-temperature cluster operator
$
  \hat T(\tau) = T^a_i(\tau)\hat \tau^a_i
  + T^{ab}_{ij}(\tau)\hat\tau^{ab}_{ij}
$
which are valid for arbitrary imaginary
times $\tau$, rather than just for the stationary case.
We insert the coupled cluster approximation for the wave function
$\exp\{\hat T(\tau)\}|\Phi\rangle$ into the imaginary time dependent \SE,
giving
\begin{equation}
  -\e^{\hat T(\tau)}\frac\partial{\partial\tau}\hat T(\tau) |\Phi\rangle =
  \hat H\e^{\hat T(\tau)} |\Phi\rangle
\end{equation}
since all excitation operators $\hat\tau^a_i,\hat\tau^{ab}_{ij},\ldots$
commute with each other. Left-multiplying the above equation with
$\exp\{-\hat T(\tau)\}$ and projecting onto excited Slater determinants
yields differential equations for the amplitudes, reading for the doubles
for instance
\begin{equation}
  -\frac\partial{\partial\tau}T^{ab}_{ij}(\tau) =
  \langle\Phi|\hat\tau^{ij}_{ab}\hat H\,\e^{\hat T(\tau)}|\Phi\rangle',
  \label{eqn:ZeroTAmplitudeDiff}
\end{equation}
where only the remaining connected terms are given, as in the case
of \Eq{eqn:ZeroTConnectedDoubles}.
Comparing \Eq{eqn:ZeroTAmplitudeDiff} and \Eq{eqn:ZeroTConnectedDoubles}
shows that the
original coupled cluster amplitudes are the steady-state solution of the
imaginary time dependent amplitudes.

\subsection{Example: direct ring coupled cluster doubles}
In order to give concrete working equations, the amplitudes can
be truncated at a certain excitation level. Truncating already
at the level of doubles still results in cumbersome equations.
We further restrict the considered contractions
to those contractions, where all fermionic loops have length 2.
The resulting theory is termed direct ring coupled cluster doubles
theory (drCCD) or, alternatively, random phase approximation (RPA) with
second order screened exchange
(SOSEX).\cite{freeman_coupled-cluster_1977,gruneis_making_2009}
Though it does not fully regard fermionic exchange relations,
it shares two important
qualities of a fully featured coupled cluster singles doubles (CCSD)
theory, where all contractions are regarded: a non-linear
nature of the amplitude equations and the ability to describe
metallic systems in the thermodynamic limit.\cite{shepherd_2013}
The calculations in Section \ref{sec:Results} are all conducted
in its linearized approximation.
For simplicity, we use the canonical set of creation and annihilation
operators, such that
$\hat H_0 = \sum_p \varepsilon_p\hat c^\dagger_p\hat c_p$.
The Hamiltonian is written in the form
$\hat H=\hat H_0 - \hat V_\mathrm{eff} + \hat V$ with the
electron-electron and the effective interaction of the reference given by
\begin{equation}
  \hat V =
    \frac12 \sum_{pqrs}V^{pq}_{sr}
    \hat c^\dagger_p \hat c^\dagger_q \hat c_r \hat c_s\,,
  \quad
  \hat V_\mathrm{eff} =
    \sum_{pq}v^p_q
    \hat c^\dagger_p \hat c_q\,,
\end{equation}
respectively. Hartree--Fock-type terms will only be considered at first
order, in accordance with RPA+SOSEX calculations in the
literature.\cite{gruneis_making_2009,harl_assessing_2010}

Integrating \Eq{eqn:ZeroTAmplitudeDiff} with the boundary condition
$\hat T(\tau\rightarrow-\infty) = 0$ and evaluating all considered contractions,
one finally obtains the drCCD equations in algebraic form,
free of any operators:
\begin{equation}
  T^{ab}_{ij}({\tau})
  = (-1)\hspace*{-2ex}\int\limits_{
    -\infty<{\tau'}<{\tau}
  }\hspace*{-2ex}
  {\d\tau'}\,
  \e^{-({\tau}-{\tau'})\Delta^{ab}_{ij}}
  \Big[
    V^{ab}_{ij}
  + V^{al}_{id}T^{db}_{lj}({\tau'})
  + V^{kb}_{cj}T^{ac}_{ik}({\tau'})
  + V^{kl}_{cd}
    T^{ac}_{ik}({\tau'})T^{db}_{lj}({\tau'})
  \Big]
  \label{eqn:ZeroTTimeAmplitudes}
\end{equation}
$\forall abij$ and implying a sum over all other indices.
In this formulation the electron-electron interaction
$V^{ab}_{ij}$ and the amplitudes $T^{ab}_{ij}$ are not antisymmetrized.
The eigenenergy difference is denoted by
$
  \Delta^{ab}_{ij}
  = \varepsilon_a + \varepsilon_b - \varepsilon_i - \varepsilon_j
$.
The drCCD amplitude equations can also be stated in terms of
(non-antisymmetrized) Goldstone diagrams
\vspace{-1ex}
\begin{equation}
  \diagramBox[1]{\includegraphics{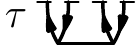}} = \\
  \hspace*{-2ex}
  \diagramBox[1]{\includegraphics{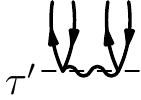}}
  + \diagramBox[1]{\includegraphics{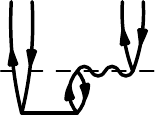}}
  + \diagramBox[1]{\includegraphics{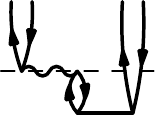}}
  + \diagramBox[1]{\includegraphics{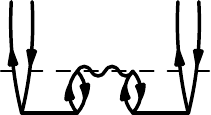}}\,,
  \label{eqn:DoublesDiagrams}
\end{equation}
depicting the contractions and the time-ordered unperturbed propagations
from $\tau'$ to $\tau$.
The imaginary time dependent formulation of the amplitude equations
is not the standard formulation, it is, however, readily generalized to
finite temperature.
Inserting the steady-state \emph{ansatz}
$T^{ab}_{ij}(\tau') = T^{ab}_{ij}(\tau) = T^{ab}_{ij}$
into \Eq{eqn:ZeroTTimeAmplitudes} and integrating over all time differences
$0<\tau-\tau'<\infty$
yields the standard form of the drCCD amplitude equations for non-degenerate
systems
\begin{equation}
  \label{eqn:ZeroTDoublesAmplitudes}
  T^{ab}_{ij} =
  \frac{
    V^{ab}_{ij}
    + V^{al}_{id}T^{db}_{lj}
    + V^{cb}_{kj}T^{ac}_{ik}
    + V^{kl}_{cd}
      T^{ac}_{ik}T^{db}_{lj}
  }{
    -\Delta^{ab}_{ij}
  }.
\end{equation}
The equations are non-linear and
contain the left-hand-side quantity $T$ in contracted forms on the
right-hand-side, which can be solved by iteration until convergence is reached.
Once the steady-state amplitudes $T^{ab}_{ij}$ are found, the
drCCD energy $E$
can be computed by considering all non-vanishing contractions in
\Eq{eqn:ZeroTConnectedEnergy}, eventually retrieving
\begin{equation}
  \label{eqn:ZeroTDoublesEnergy}
  E = \varepsilon_i - v^i_i\\
  + \frac12\left(V^{ij}_{ij} - V^{ij}_{ji}\right)
  + \frac12\left(V^{ij}_{ab}-V^{ji}_{ab}\right)T^{ab}_{ij}
\end{equation}
implying a sum over all indices.
The term involving the amplitudes
is referred to as drCCD correlation energy, the remaining term
is the Hartree-exchange energy $E_\mathrm{HX}$.

\section{Finite temperature coupled cluster}
\label{sec:FiniteTemperature}
The connected and fully contracted expectation value giving the
zero temperature coupled cluster energy in \Eq{eqn:ZeroTConnectedEnergy}
corresponds to a subset of the zero temperature many-body perturbation
expansion\footnote{The
expansion is only considered formally to identify
the function returning the energy from the electron repulsion integrals,
the eigenenergies, and the ensemble parameters. Individual terms may diverge.}.
The truncated amplitude equations in the Dyson-like form, as for instance in
\Eq{eqn:DoublesDiagrams}, define a recursive rule
how to generate terms of the perturbation expansion
by connecting sub-terms with a finite maximum number of open connections
to a more complex sub-term with the same finite maximum number of open
connections. The sub-terms are connected by the application of the
perturbation.
CCSD, for instance, considers all possible connections of the
sub-terms with each other having at most four open connections.
All open connections are going upwards, so the perturbations
occur in (imaginary) time ordered fashion. The number of connections
is always even and they come in particle/hole pairs since the
perturbation preserves the number of electrons.
The sub-terms are connected in topologically distinct ways, such
that a single MBPT term, here denoted by a Goldstone diagram,
will be evaluated exactly once or not at all.
Truncating the cluster operator at higher and higher levels considers
more and more open connections of the sub-terms, approaching the
full many-body perturbation expansion.

One can therefore generalize coupled cluster to finite temperature
by translating the perturbation expansion from zero temperature
to finite temperature, as outlined in \Fig{fig:FiniteTempMBPT}.
At finite temperature the imaginary times of each occurring perturbation
are integrated over the finite domain $[0,\beta]$, rather than over
the infinite domain of the zero-temperature case. The last perturbation
at $\tau_n$ also needs to be integrated, rather than being kept fixed.
Furthermore, the contractions within quantum
mechanical expectation values of the Hartree--Fock or DFT Slater determinant
$
\langle\Phi|
  \ldots\contraction{}{\hat c}{_a}{\hat c}
  \hat c_a \hat c^\dagger_b\ldots
|\Phi\rangle_0
$
are superseded by contractions within the ensemble average of the
respective finite-temperature mixture.
Each single-particle
level of the Hartree--Fock or DFT calculation must be allowed to
exchange its electron with the bath if the occupation probabilities
are to be independent of each other, enforcing the grand canonical ensemble.
In the canonical basis the (non-normalized) density operator factorizes into
\begin{equation}
  \hat\rho_0 = \e^{-\beta(\hat H_0 - \mu\hat N)}
  = \bigotimes_p \e^{-\beta(\varepsilon_p-\mu)\hat n_p}
\end{equation}
with $\hat n_p=\hat c^\dagger_p\hat c_p$ and
the fixed chemical potential of the environment $\mu$.
The probability of occupation $f_p$ of a single
level $p$ is then given by the Fermi--Dirac distribution
\begin{equation}
  \label{eqn:fp}
  f_p = \langle \hat n_p \rangle_0
  = \frac{\Tr\{\hat\rho_0\hat n_p\}}{\Tr\{\hat\rho_0\}}= \frac{
     \e^{-\beta(\varepsilon_p-\mu)}
  } {
    1+\e^{-\beta(\varepsilon_p-\mu)}
  }
\end{equation}
and the probability of vacancy of the level $p$ is denoted by $f^p = 1-f_p$.
Matsubara generalized Wick's theorem to
mixtures\cite{matsubara_new_1955,thouless_2014} allowing contractions to
be defined for a thermal Hartree--Fock or DFT reference:
\begin{equation}
  \langle
    \ldots\contraction{}{\hat c}{^\dagger_i}{\hat c}
    \hat c^\dagger_i \hat c_j\ldots
  \rangle_0 = \delta^i_j f_i\,, \quad
  \langle
    \ldots\contraction{}{\hat c}{_a}{\hat c}
    \hat c_a \hat c^\dagger_b\ldots
  \rangle_0 = \delta^b_a f^a.
\end{equation}
Products of occupation or vacancy probabilities are denoted by
$f^{ab\ldots}_{ij\ldots}=f^af_if^bf_j\ldots$, where lower and upper indices
are referred to as thermal hole and thermal particle levels,
although all indices $i,a,\ldots$ are to be summed, in principle,
over all single-body levels $p$.
In practice, however, the sum over thermal holes $i$ and thermal particles $a$
can be restricted to levels where $f_i$ and $f^a$ is non-negligible,
respectively.
The fact that thermal hole indices $i$ and thermal particle indices $a$
can refer to the same level $p$
spawns contractions which are not present in the usual zero-temperature
formalism, in which hole and particle states are disjoint.
This lays the fundament for finite temperature many-body perturbation theory,
in fact, including Hartree--Fock itself, defining the exchange free energy.
Appendix \ref{sec:Derivation} lists a brief derivation of finite
temperature many-body perturbation theory. Finite temperature Hartree--Fock
and DFT are discussed in.\cite{mermin_1963,mermin_1965,eschrig_2010}

In the case where the Hartree--Fock or the DFT reference is degenerate
in the zero-temperature limit, diagrams with states that are holes as well
as particles remain with non-vanishing contributions.
In the degenerate case the zero-temperature limit of the grand canonical
ensemble at constant chemical potential does not agree with the
zero-temperature theory with a fixed number of electrons.
This is referred to as the Kohn--Luttinger
conundrum\cite{kohn_ground-state_1960} and it is treated in detail in
Santra and Schirmer.\cite{santra_2017}

\subsection{Finite temperature coupled cluster equations}
Translating the MBPT terms means working with the fully contracted
operators, rather than with the operators itself. Rather than
defining a cluster operator $\hat T$, we define how operators must
occur within the fully contracted expectation values of an MBPT term.
They are required to have both contractions going upwards,
towards positive times:
\begin{equation}
  \label{eqn:FTCluster}
  \langle\
  \contraction[2ex]{}{(\,\cdot\,)}{\ T^a_i
    \hat c^\dagger_a}{\hat c}
  \contraction{}{(\,\cdot\,)}{\ T^a_i}
    {\hat c}
  (\,\cdot\,)\ T^a_i
    \hat c^\dagger_a\hat c_i\
  \rangle_0'
  = \diagramBox[0.0]{\includegraphics{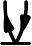}},
  \quad
  \langle\
    \contraction[2ex]{}{(\,\cdot\,)}{\ T^{ab}_{ij}
    \hat c^\dagger_a\hat c^\dagger_b \hat c_j}{\hat c}
    \contraction{}{(\,\cdot\,)}{\ T^{ab}_{ij}}
    {\hat c}
    \bcontraction[2ex]{}{(\,\cdot\,)}{\ T^{ab}_{ij}
    \hat c^\dagger_a\hat c^\dagger_b} {\hat c}
    \bcontraction{}{(\,\cdot\,)}{\ T^{ab}_{ij}
    \hat c^\dagger_a}{\hat c}
    (\,\cdot\,)\ T^{ab}_{ij}
    \hat c^\dagger_a\hat c^\dagger_b \hat c_j\hat c_i\
  \rangle_0'
   = \diagramBox[0.0]{\includegraphics{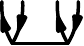}}.
\end{equation}
Some
contractions may no longer vanish at finite temperature, so all of them
have to be reconsidered.
The algebra of contractions is, however, identical, most importantly
\begin{equation}
  \langle\
    \contraction[2ex]{}{(\,\cdot\,)}{\
      \hat c^\dagger_a \hat c_i \ \hat c^\dagger_b}{\hat c}
    \contraction{}{(\,\cdot\,)}{\
      \hat c^\dagger_a \hat c_i \ }{\hat c}
    \bcontraction[2ex]{}{(\,\cdot\,)}{\
      \hat c^\dagger_a}{\hat c}
    \bcontraction{}{(\,\cdot\,)}{\ }{\hat c}
    (\,\cdot\,)\
      \hat c^\dagger_a \hat c_i \ \hat c^\dagger_b\hat c_j\
  \rangle_0'
  =
  +\langle\
    \bcontraction[2ex]{}{(\,\cdot\,)}{\
      \hat c^\dagger_b \hat c_j \ \hat c^\dagger_a}{\hat c}
    \bcontraction{}{(\,\cdot\,)}{\
      \hat c^\dagger_b \hat c_j \ }{\hat c}
    \contraction[2ex]{}{(\,\cdot\,)}{\
      \hat c^\dagger_b}{\hat c}
    \contraction{}{(\,\cdot\,)}{\ }{\hat c}
    (\,\cdot\,)\
      \hat c^\dagger_b \hat c_j \ \hat c^\dagger_a\hat c_i\
  \rangle_0'\,.
\end{equation}
Thus, the differential equations for the amplitude scalars, as for instance
\Eq{eqn:ZeroTAmplitudeDiff}, also hold in the finite temperature case.
The equation for the doubles amplitudes reads
for instance
\begin{equation}
  -\frac\partial{\partial\tau}T^{ab}_{ij}(\tau) =
  \langle
    \contraction[2ex]{}{\hat c}{^\dagger_i\hat c^\dagger_j \hat c_b\hat c_a
    }{(\hat H\,\e^{\hat T(\tau)})}
    \contraction{\hat c^\dagger_i\hat c^\dagger_j \hat c_b}{\hat c}{_a
    }{(\hat H\,\e^{\hat T(\tau)})}
    \bcontraction[2ex]{\hat c^\dagger_i}{\hat c}{^\dagger_j \hat c_b\hat c_a
    }{(\hat H\,\e^{\hat T(\tau)})}
    \bcontraction{\hat c^\dagger_i\hat c^\dagger_j} {\hat c}{_b\hat c_a
    }{(\hat H\,\e^{\hat T(\tau)})}
    \hat c^\dagger_i\hat c^\dagger_j \hat c_b\hat c_a
    (\hat H\,\e^{\hat T(\tau)})
  \rangle_0',
  \label{eqn:AmplitudeDiff}
\end{equation}
where no contractions among the $\hat T$s are allowed according to
the requirement stated in \Eq{eqn:FTCluster}.
Hartree--Fock-type contractions of $\hat H_1$ with itself are, however, allowed.

Having solved the imaginary time dependent amplitudes for all times
$\tau$ between $0$ and $\beta$, the coupled cluster grand potential
can be computed from the grand potential of the reference $\Omega_0$
and a reference thermal expectation value of all connected and fully
contracted coupled cluster terms with the perturbation part $\hat H_{1}$
\begin{equation}
  \Omega = \Omega_0 -\frac1\beta
  \int_0^\beta\d\tau\,\langle
  \contraction{}{\hat H_{1}}{\,}{(\e^{\hat T(\tau)})}
  \hat H_{1}\,(\e^{\hat T(\tau)})
  \rangle_0'\,,
\end{equation}
according to \Eq{eqn:FreeEnergyDifference}. Also here, contractions
of $\hat H_1$ with itself are allowed, contractions among $\hat T$s are not.
Note that in the grand canonical ensemble the expectation value of the number
operator $\langle\hat N\rangle$ will be affected by correlation.
In principle, the chemical potential $\mu$ needs to be scanned for
the value $\mu_N$, such that $\langle\hat N\rangle$ is fixed to the
desired number of electrons $N$:
\begin{equation}
  \left.\frac{\partial\log\mathcal Z(\beta,\mu)}{\beta\partial\mu}\right|_
  {\mu=\mu_N}
  = N.
\end{equation}
The differences $\mu_N-\mu_{N-1}$ and $\mu_{N+1}-\mu_N$ correspond to the
finite-temperature generalization of the ionization potential (IP) and the
electron affinity (AE), respectively.

\subsection{Example: direct ring coupled cluster doubles}
The concrete working equations of a zero temperature coupled cluster theory,
formulated in imaginary time as in \Eq{eqn:ZeroTTimeAmplitudes},
can be translated to finite temperature by
\begin{itemize}
\item extending hole and particle states to overlapping thermal hole and
  particle states,
\item convolving with the time evolution operator
  $\e^{-(\tau-\tau'){\Delta}^{ab\ldots}_{ij\ldots}}$
  over the finite domain $[0,\tau]$, and
\item multiplying with the thermal occupancies $f^{c\ldots}_{k\ldots}$ for
  each contracted thermal particle/hole index.
\end{itemize}
Applied to the direct ring coupled cluster doubles theory, we obtain
\begin{multline}
  T^{ab}_{ij}({\tau})
  = (-1)\int_0^\tau \d\tau'\,
  \e^{-({\tau}-{\tau'}){\Delta}^{ab}_{ij}}
  \Big[
    V^{ab}_{ij}
  + f^d_l V^{al}_{id}T^{db}_{lj}({\tau'}) \\
  + f^c_k V^{kb}_{cj}T^{ac}_{ik}({\tau'})
  + f^{cd}_{kl} V^{kl}_{cd}
    T^{ac}_{ik}({\tau'})T^{db}_{lj}({\tau'}) \Big]
  \label{eqn:drCCDTimeDoubles}
\end{multline}
for the doubles amplitudes, implying a sum over all indices occurring
only on the right-hand-side.
Having solved for the imaginary time dependent amplitudes, the
direct ring coupled cluster doubles grand potential is evaluated by
\begin{equation}
  \Omega = \Omega_\mathrm{HX}
  +\diagramBox{\includegraphics{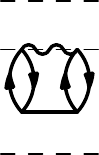}}
  +\diagramBox{\includegraphics{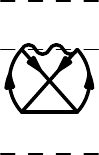}}
  = \Omega_\mathrm{HX}
  + \frac1\beta\int_0^\beta {\d\tau}\,\Big[
    \frac12f^{ab}_{ij} \left(V^{ij}_{ab}-V^{ji}_{ab}\right)
    T^{ab}_{ij}(\tau)\Big]\,.
  \label{eqn:DoublesEnergy}
\end{equation}
$\Omega_\mathrm{HX}$ denotes the DFT Hartree--Exchange
grand potential
\begin{equation}
  \Omega_\mathrm{HX} =
  f_i(\varepsilon_i-\mu) + \frac1\beta (f_i\log f_i+f^i\log f^i)
  - f_i v^i_i + \frac12 f_{ij}\left(V^{ij}_{ij} - V^{ij}_{ji}\right).
\end{equation}

\section{Solving the finite temperature amplitude equations}
\label{sec:Solving}
At zero temperature only the steady-state solution of the amplitudes
is required, while in the finite-temperature case the free energy
is determined from averaging over all imaginary times $[0,\beta]$.
In the finite-temperature case one indeed needs to solve the system of
coupled integral equations
(\ref{eqn:drCCDTimeDoubles}) to
an extent permitting sufficient accuracy in the quadrature of
\Eq{eqn:DoublesEnergy}.
The system of equations is non-linear and can thus only be solved
iteratively, representing the amplitudes on
an imaginary time grid, which is not necessarily equidistant.
Which choice of the grid points
provides sufficient accuracy depends on the system and on the
temperature $1/\beta$ of the calculation.
The amplitude integral equations can be solved on the grid one-by-one, starting
at $T^{ab}_{ij}(\tau_0=0) = 0$ and with a guess for $T^{ab}_{ij}(\tau_1)$
at the end of the first interval.
From amplitudes at the beginning and at the end of the current interval
a continuous interpolation to arbitrary imaginary times $T(\tau')$
can be constructed and
inserted into the right-hand-side of the amplitude equations to be solved,
e.g.~\Eq{eqn:drCCDTimeDoubles}.
Integrating the amplitude equations from the beginning to the end of the
interval, employing the
interpolation on the r.h.s., gives an updated guess for the amplitudes
at the end of the interval $\tilde T(\tau_1)$,
which should agree with the initial amplitudes $T(\tau_1)$.
From the deviation of $\tilde T(\tau_1)$ from $T(\tau_1)$
new amplitudes are estimated at $\tau_1$.
Once convergence is reached the next interval can be solved.
This procedure only requires storing the amplitudes at the beginning
and the end of the current interval and convergence accelerating techniques,
such as direct inversion of iterative subspace (DIIS) can be
employed.\cite{pulay_1980,pulay_1982,scuseria_1986}
However, the overall accuracy depends strongly on the choice of grid points
and the amplitudes must be well-converged at each grid point to minimize
error accumulation.

\subsection{Linearized direct ring coupled cluster}
Neglecting quadratic and higher order terms of the coupled
cluster amplitude equations leaves a linear, though inhomogeneous system of
equations, which can be solved by diagonalization.
Once obtained, the imaginary time dependence of the
amplitudes can be evaluated and integrated analytically.
In the particular case of the linearized direct ring approximation
the matrix is small enough
such that diagonalization is computationally feasible.
We will therefore employ this approximation to study temperatures as low
as room temperature.

The linearized direct ring coupled cluster amplitude equations
are given by
\begin{multline}
  T^{ab}_{ij}(\tau) =
  \diagramBox[1]{\includegraphics{T}} =
  \hspace*{-2ex}
  \diagramBox[1]{\includegraphics{V}}
  + \diagramBox[1]{\includegraphics{TV}}
  + \diagramBox[1]{\includegraphics{VT}} \\
  = (-1)\int_0^\tau \d\tau'\,
  \e^{-({\tau}-{\tau'}){\Delta}^{ab}_{ij}}
  \Big[ V^{ab}_{ij}
  + f^c_k V^{ak}_{ic} T^{cb}_{kj}(\tau')
  + f^d_l V^{lb}_{dj} T^{ad}_{il}(\tau')
  \Big]
\end{multline}
Inserting the amplitudes into \Eq{eqn:DoublesEnergy}
gives the expansion of the grand potential in the
linearized drCCD approximation.
In this approximation the left and the right particle/hole pairs do not
interact and their propagation can be factorized. A particle/hole pair in the
states $c,k$ at $\tau_1$ can propagate independently of the other pair into
the states $b,j$ at the later time $\tau_2$. The propagation
can be either free or via one or more electron-electron interactions.
It can be given by the following Dyson-like equation
\begin{multline}
  G^{bk}_{jc}(\tau_2,\tau_1) = \diagramBox{\includegraphics{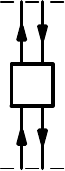}}
  = \diagramBox{\includegraphics{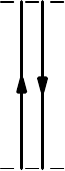}}
  + \diagramBox{\includegraphics{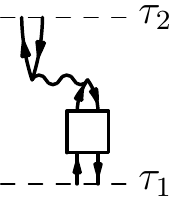}} \\
  = \delta^b_c\delta^k_j \e^{-(\tau_2-\tau_1)\Delta^b_j}
   -\int_{\tau_1}^{\tau_2} \d\tau'\,
    \e^{-(\tau_2-\tau')\Delta^b_j}
    g^{bd}_{jl} V^{bl}_{jd} G^{dk}_{lc}(\tau',\tau_1)\,,
  \label{eqn:SinglesDyson}
\end{multline}
where $g^{bd}_{jl}=(f^{bd}_{jl})^{1/2}$ denotes the square roots
of the Wick contraction weights. The square roots of the contraction weights
are assigned to each leg of the electron-electron interaction
in order to retain a symmetric form of the propagator, satisfying
\begin{equation}
  \int_{\tau_1}^{\tau_2} \d\tau'\,
    \e^{-(\tau_2-\tau')\Delta^b_j}
    g^{bd}_{jl} V^{bl}_{jd} G^{dk}_{lc}(\tau',\tau_1) \\
  = \int_{\tau_1}^{\tau_2} \d\tau'\,
    G^{bl}_{jd}(\tau_2,\tau')
    g^{cd}_{kl} V^{dk}_{lc}
    \e^{-(\tau'-\tau_1)\Delta^c_k}
\,.
\end{equation}
Identifying the left two indices of $G$ in \Eq{eqn:SinglesDyson}
as the row index, and the right two indices as the column of a matrix
$\vec G$, the particle/hole propagation is given by the matrix exponent
\begin{equation}
  \vec{G}(\tau_2,\tau_1) = \exp\{-(\tau_2-\tau_1)\vec A\}
\end{equation}
with the effective particle/hole interaction
\begin{equation}
  A^{bk}_{jc}=
  \delta^b_c\delta^k_j(\varepsilon_b-\varepsilon_j) + g^{bc}_{jk} V^{bk}_{jc}.
  \label{eqn:TdaA}
\end{equation}
This corresponds to a generalization of the Casida equation in the
Tamm--Dancoff approximation to finite
temperature.\cite{dreuw_2005,scuseria_2008}
The matrix $\vec A$ is hermitian permitting an eigendecomposition
with real eigenvalues
\begin{equation}
  A^{bk}_{jc} =
    \delta^b_c\delta^k_j\Delta^b_j
  + g^{bc}_{jk} V^{bk}_{jc}
  = U^{b}_{jF}\Lambda_F^F{U^\ast}^{cF}_k,
\end{equation}
implying a sum over the eigenvalue index $F$.
The particle/hole imaginary time evolution operator is then given
by
\begin{equation}
  G^{bk}_{jc}(\tau_2,\tau_1)
  = U^{b}_{jF}\e^{-(\tau_2-\tau_1)\Lambda_F^F}{U^\ast}^{cF}_k.
\end{equation}
Finding the eigenvalue decomposition scales as
$\mathcal O(N_\mathrm{v}^3N_\mathrm{o}^3)$ with the number of thermal holes
$N_\mathrm{o}$ and particles $N_\mathrm{v}$, which is of the same order
as zero temperature coupled cluster doubles.

Having found the eigenvalue decomposition, one can
evaluate the linearized direct ring coupled cluster doubles amplitudes
at the time $\tau$ by propagating the left and the right particle/hole
pairs from the initial electron-electron interaction $V^{cd}_{kl}$ at $\tau'$
to the time $\tau$
\begin{equation}
  T^{ab}_{ij}(\tau) = (-1) \int_0^\tau\d\tau'\,
  G^{ak}_{ic}(\tau,\tau')\,G^{bl}_{jd}(\tau,\tau')\,g^{cd}_{kl}V^{cd}_{kl}.
  \label{eqn:TdaE}
\end{equation}
The correlation contribution to the grand potential
in the linearized direct ring coupled cluster doubles
approximation is then found to be
\begin{multline}
  \Omega - \Omega_\mathrm{HX}
  =  \diagramBox{\includegraphics{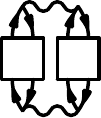}}
   + \diagramBox{\includegraphics{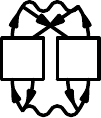}}
  =
  \frac1\beta \int_0^\beta\d\tau\,
  \frac12\, g^{ab}_{ij}
  \left(V^{ij}_{ab}-V^{ji}_{ab}\right) T^{ab}_{ij}(\tau) \\
  =
  -\left(\frac1{\Lambda^{FG}} +
    \frac{\e^{-\beta\Lambda^{FG}} - 1}{\beta(\Lambda^{FG})^2}
  \right)\,\frac12\,\left(W_{FG}-X_{FG}\right)\,W^{FG}
  \label{eqn:TDA}
\end{multline}
with $\Lambda^{FG} = \Lambda^F_F+\Lambda^G_G$ and where $W$ and $X$
denote the direct and the exchange electron-electron interaction
in the eigenmodes of the particle/hole propagator, respectively
\begin{equation}
  W_{FG} = U_{kF}^cU_{lG}^d g^{cd}_{kl} V_{cd}^{kl}, \quad
  X_{FG} = U_{kF}^cU_{lG}^d g^{cd}_{lk} V_{cd}^{lk}.
\end{equation}
A sum over all repeated indices on the right-hand-side is implied
and $W^{FG} = \overline{W_{FG}}$.
Note that the initial and the final interaction also need to carry the
square root of the contraction weights.
Computing other observables than the grand potential requires derivatives
of the log-partition function with respect to $\beta$ or $\mu$,
according to Appendix \ref{ssc:Observables}.
They have to be computed numerically, as the square roots of the Fermi
weights occur in the effective particle/hole interaction
$\vec A$ to be diagonalized.

For finite $\beta$ the above expression will be finite, also for terms
with vanishing eigenvalues.
However, degenerate states having non-vanishing probability of
both, occupancy as well as of vacancy, need to be treated specially.
For such states $p,q,r,s$ (being not necessarily all distinct)
the effective particle/hole interaction $A^{pq}_{sr}$ of \Eq{eqn:TdaA}
will be proportional
to the electron-electron interaction
$V^{pq}_{sr}$ since all eigenenergies are the same
and $g^{pr}_{sq}=g$ is constant.
For $p\neq s$, two distinct rows $(p,s)\neq(s,p)$ in the matrix
$\vec A$ will be identical in the case of real-valued orbitals:
\begin{equation}
  A^{pq}_{sr} = g\,V^{pq}_{sr} = g\,V^{sq}_{pr} = A^{sq}_{pr}.
\end{equation}
This will render $N_\mathrm{d}(N_\mathrm{d}-1)/2$ eigenvalues of $\vec A$ zero
for a group of $N_\mathrm{d}$ degenerate states serving equally
as particles and as holes. This means that the amplitudes have
more degrees of freedom than there is information taken into account
for their propagation.
However, in the case of real-valued orbitals the matrix of the
initial electron-electron interaction $g\,V^{sq}_{qr}$ in \Eq{eqn:TdaE}
is also identical to that of the particle/hole effective interaction
$A^{pq}_{sr}$ for degenerate states $p,q,r,s$.
Therefore, eigenmodes $F$ of $\vec A$ with an eigenvalue of zero
(purely within the degenerate space) also also have zero coupling to the
initial electron-electron interaction and can be disregarded.

\subsection{Zero temperature limit}
\label{ssc:ZeroTempLimit}
The limit of vanishing temperature or infinite $\beta$ requires to
distinguish between cases where the unperturbed reference is
non-degenerate and where it is degenerate.
In the non-degenerate case, the matrix $\vec A$ will be positive definite,
having only positive eigenvalues $\Lambda^F_F$.
The correlation grand potential in the linearized drCCD of \Eq{eqn:TDA}
can thus be evaluated in the limit $\beta\rightarrow\infty$
arriving at the zero-temperature expression for the linearized direct ring
coupled cluster doubles energy, evaluated from solving the Casida
equation in the Tamm--Dancoff approximation\cite{scuseria_2008}
\begin{equation}
  \Omega - \Omega_\mathrm{HX} =
  -\frac12\,\frac{
    \left(W_{FG}-X_{FG}\right)\,W^{FG}
  }{
    \Lambda^{FG}
  }.
\end{equation}
In the presence of a finite gap of the occupied and the virtual states
there will be no correlation contribution to the expectation
value of the number operator since $\mu$ can be varied without
affecting the correlation contribution to the log-partition function.
Thus, the above equation will also be the correlation energy of
the zero temperature linearized direct ring coupled cluster
doubles approximation.

In the case of $N_\mathrm{d}$ degenerate states at the Fermi level
in the limit of $T\rightarrow0$,
the matrix $\vec A$ of the effective particle/hole
interaction exhibits $N_\mathrm{d}(N_\mathrm{d}-1)/2$ eigenmodes with an
eigenvalue of zero.
As discussed in the previous section,
the initial electron-electron interaction does not couple to these modes and
the ldrCCD correlation grand potential also yields a finite value in the
degenerate case for $\beta\rightarrow\infty$.
However, $\mu$ cannot be varied without changing the correlation contribution
to the log-partition function. Thus, the zero-temperature limit of the
correlation grand potential at the given $\mu$ will not yield a
zero-temperature correlation energy.

\section{Applications}
\label{sec:Results}
\begin{figure*}[t]
\begin{center}
  \includegraphics{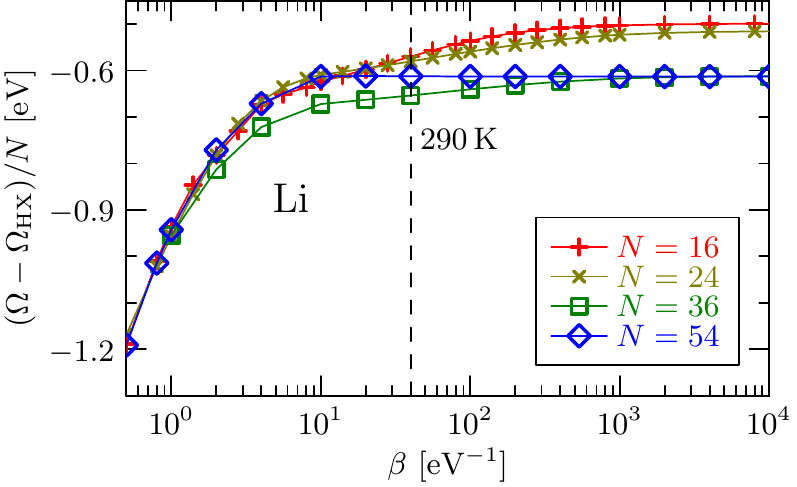}
  \hfill
  \includegraphics{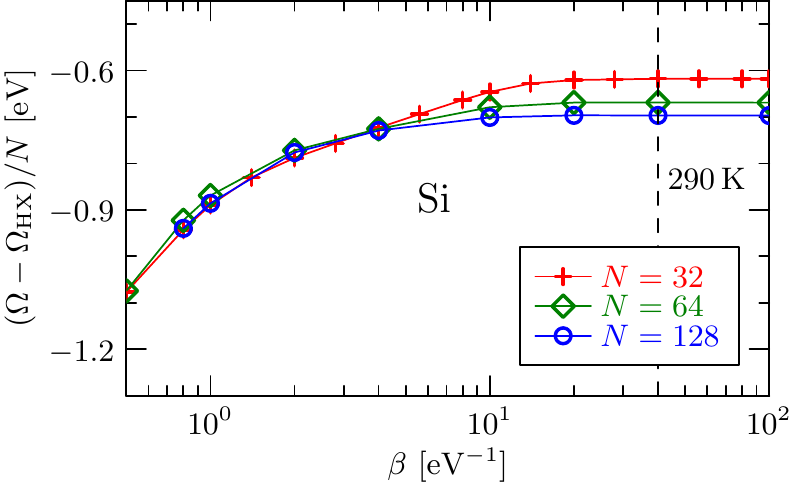}
\end{center}
\caption{
  Correlation contribution per electron to the grand potential in the
  linearized direct ring coupled cluster doubles
  approximation for solid lithium (Li) and solid silicon (Si)
  for various system sizes
  as a function of the inverse temperature $\beta$.
  In the
  insulating case the correlation grand potential approaches the ground state
  correlation energy obtained from the zero-temperature theory.
  In the metallic case the correlation grand potential converges to different
  values not only due to finite-size errors. Different types of degeneracies
  at low temperature are expected to give different correlation contributions
  to the
  expectation value of the number operator. Only the system with 54 electrons
  is non-degenerate at low temperatures.
}
\label{fig:LiSi}
\end{figure*}

In this section the presented finite temperature coupled cluster framework
is applied to solid lithium, a metallic system, and to solid silicon as
an insulating system, to demonstrate its practical applicability. 
All calculations were conducted with the Coupled Cluster for Solids
\texttt{cc4s} code, developed at TU Wien, based on a
density functional theory reference, provided by the Vienna \emph{ab-initio}
Simulation Package
(VASP)\cite{blochl_projector_1994,kresse_norm-conserving_1994,kresse_efficient_1996}.
The linearized direct ring coupled cluster doubles approximation was chosen
as a test as it is expected to converge for metallic systems
at zero temperature in the thermodynamic limit of an infinite solid.
The calculations are not fully converged
with respect to the number of virtual orbitals and with respect
to the thermodynamic limit. However, the systems provide the respective
key features of interest. For most system sizes
the lithium system is truly degenerate at all
temperatures considered.
The silicon system provides an underestimated but finite
band gap and can therefore also be calculated with the conventional zero
temperature linearized direct ring coupled cluster doubles methods,
also known as TDA+SOSEX.

The considered lithium super-cells comprise 16, 24, 36, and 54 electrons in
approximately 5 restricted orbitals per atom at the $\Gamma$-point.
The silicon super-cells contain 8, 16, and 32 atoms in approximately
16 restricted orbitals per atom at the $\Gamma$-point.
Only the valence electrons were contained in the calculations, freezing
the He and the Ne core of Li and Si, respectively.
All super-cells were relaxed at the $k$-point-converged DFT level
with the Perdew--Burke--Enzerhof (PBE) functional optimized for solids. The
kinetic cutoff energy for the calculation of the electron-electron
interaction was set to 250\,eV.
For large systems the diagonalization of the particle/hole effective
interaction, given in \Eq{eqn:TdaA}, poses the most demanding part
of the calculation, scaling as $\mathcal O(N_\mathrm{v}^3N_\mathrm{o}^3)$.
The matrix dimensions of the largest considered systems
were about $100000\times100000$ for high temperatures and
$25000\times25000$ for low temperatures, which is still feasible for
standard diagonalization routines, as implemented in \texttt{ScaLAPACK}.

\Fig{fig:LiSi} plots the resulting linearized direct ring
coupled cluster doubles (ldrCCD) correlation grand potentials
as a function of the inverse temperature $\beta$.
In the case of the semi-conducting Si, the finite-temperature correlation
grand potential approaches the correlation energy, calculated with the existing
zero-temperature formalism for the respective system size,
as expected. In the case of the metallic Li,
the calculations of different sizes converge to different values for low
temperatures not only due to finite-size errors. For low temperatures
the systems of 16, 24, and 36 atoms exhibit different types of degeneracies
at the Fermi edge having 2 electrons in 6 orbitals, 2 electrons in 8 orbitals,
and 14 electrons in 8 orbitals.
The system of 54 atoms is non-degenerate at low temperatures.
The different types of degeneracies lead to different correlation
contributions to the grand potential.
The correlation contributions to the expectation
value of the number operator $\langle\hat N\rangle$ are also expected
to be different.
Twist averaging\cite{filippi_1999,lin_2001,holzmann_2016}
is expected to improve on the finite-size convergence,
however, its implementation into \texttt{cc4s} for metallic systems goes
beyond the scope of this work.

\section{Summary}
This work presents a framework for finite temperature coupled cluster
theories, including a practical algorithm to apply it to extended systems.
In this framework, coupled cluster theories are viewed as rules
to generate an (infinite) subset of the finite temperature many-body
perturbation expansion.
When increasing the truncation level of the cluster operator the
subset becomes more and more complete.

For demonstration purposes linearized direct ring coupled cluster doubles
correlation grand potentials are calculated for two solids,
metallic lithium and semi-conducting silicon.
The chosen approximation allows an exact evaluation of the imaginary time
behavior by matrix diagonalization for arbitrarily low temperatures,
also for degenerate systems.
For all sizes of silicon and for the 54 atom super-cell of lithium
the low temperature DFT reference is non-degenerate.
The low- but finite-temperature results for these systems
can thus be compared with existing
non-degenerate zero temperature linearized drCCD results and agreement is found.

Going beyond linearized drCCD requires iterative schemes and a sampling
of the imaginary time dependent amplitudes. At high enough temperatures
full coupled cluster singles and doubles converges on a uniform grid,
as done by White and Chan.\cite{white_2018}
Which schemes together
with which non-linear approximations allow to go to low temperatures
will be subject of future investigation. Certainly, excitations
of single electrons, rather than of doubly occupied orbitals,
should not be neglected in finite-temperature theories beyond
direct ring coupled cluster doubles.
Finally, it is worth to remark that
derivatives with respect to $\beta=1/k_\mathrm{B}T$
give access to central statistical moments of the
\emph{interacting} Hamiltonian, exhibiting properties of the density
of states.
It remains to be studied how the proposed finite temperature
coupled cluster (FT-CC) framework compares to thermalizing the correlated
excited states, obtained by an equation of motion coupled cluster (EOM-CC)
calculation\cite{monkhorst_1977,stanton_1993}.
The two approaches work fundamentally different:
EOM-CC describes correlated excited states by a discrete set of amplitudes
for each excited state, while in FT-CC correlation is described
by continuous amplitude functions.

\section*{Acknowledgements}
The author thanks the reviewers for pointing out an error in the
original derivation.
Fruitful discussions with
Andreas Gr\"uneis, Andreas Irmler and Joachim Burgd\"orfer
are gratefully acknowledged.

\appendix
\section{Brief derivation of finite temperature many-body perturbation theory}
\label{sec:Derivation}
This appendix summarizes finite temperature MBPT following the notation
of Matsubara's original work \cite{matsubara_new_1955}. It can be found in
more detail, for instance in \cite{thouless_2014}.

In thermal equilibrium, the (non-normalized) density operator $\hat\rho$
of the grand canonical ensemble with chemical potential $\mu$ is given by
\begin{equation}
  \hat\rho = {\rm e}^{-\beta(\hat H-\mu\hat N)},
  \label{eqn:DensityOperator}
\end{equation}
where $\beta=1/k_{\rm B}T$ is the inverse temperature.
Applying the Zassenhaus formula separates the density operator
into a reference part $\hat\rho_0$ and a correlation part $\hat S$:
\begin{equation}
  \hat\rho = {\rm e}^{-\beta(\hat H_0+\hat H_1-\mu\hat N)}
  = \underbrace{{\rm e}^{-\beta(\hat H_0-\mu\hat N)}}_{\hat\rho_0}
    \underbrace{
      {\rm e}^{-\beta\hat H_1}
      {\rm e}^{+\beta^2/2[\hat H_0,\hat H_1]}
      \ldots
    }_{\hat S}
  \label{eqn:CorrelatedDensity}
\end{equation}
since $\hat N$ commutes with $\hat H_0$ and $\hat H_1$.

\subsection{Bloch equation}
The derivative of $\hat\rho$ with respect to $\beta$ defines an equation of
motion for $\hat\rho$ as a function of $\beta$, referred to as Bloch equation
\begin{equation}
  -\frac{\partial\hat\rho}{\partial\beta} = (\hat H-\mu\hat N)\hat\rho\,.
\end{equation}
Inserting Eq.~\eqref{eqn:CorrelatedDensity} on both sides and
using $-\partial\hat\rho_0/\partial\beta = (\hat H_0-\mu\hat N)\hat\rho_0$
yields
\begin{equation}
  -\frac{\partial(\hat\rho_0\hat S)}{\partial\beta}
  = (\hat H_0-\mu\hat N)\hat\rho_0\hat S -
    \hat\rho_0 \frac{\partial\hat S}{\partial\beta}
  = (\hat H_0+\hat H_1-\mu\hat N)\hat\rho_0\hat S\,,
\end{equation}
from which the equation of motion for the correlated
part $\hat S$ follows
\begin{equation}
  -\frac{\partial\hat S}{\partial\beta} = \hat H_1(\beta)\hat S(\beta)\,,
  \label{eqn:CorrelatedDensityEOM}
\end{equation}
with $
  \hat H_1(\tau)=
    {\rm e}^{+\tau(\hat H_0-\mu\hat N)}
    \hat H_1
    {\rm e}^{-\tau(\hat H_0-\mu\hat N)}
$.
This is the finite-temperature analog to the interaction picture.

\subsection{Perturbation expansion of $\hat S$}
The equation of motion for the correlation part $\hat S$
can be transformed into a Voltera integral equation with the formal solution
\begin{equation}
  \hat S(\beta) = \sum_{n=0}^\infty{(-1)}^n
     \int\limits_{0<\tau_1<\ldots<\tau_n<\beta}
    {\rm d}\tau_1\ldots{\rm d}\tau_n
    \hat H_1(\tau_n) \ldots \hat H_1(\tau_1)\,.
  \label{eqn:SExpansion}
\end{equation}
Note that $\hat S(\beta)$ also depends on $\mu$ although this is not explicitly
denoted.

\subsection{Grand potential difference}
The grand canonical partition functions of the fully interacting system
$\mathcal Z$ and of the reference system $\mathcal Z_0$ are given by
\begin{equation}
  \mathcal Z(\beta,\mu) = \Tr\{\hat\rho_0\hat S(\beta)\}, \quad
  \mathcal Z_0(\beta,\mu) = \Tr\{\hat\rho_0\}
\end{equation}
respectively. We are interested in the grand potential difference between
the fully interacting and the reference system
\begin{equation}
  \Delta \Omega=-\frac1\beta\big(
    \log\mathcal Z(\beta,\mu) - \log\mathcal Z_0(\beta,\mu)
  \big)
  =-\frac1\beta \log\big\langle\hat S(\beta)\big\rangle_0\,,
\end{equation}
where
$\langle\hat A\rangle_0=\Tr\{\hat\rho_0\hat A\}/\mathcal Z_0(\beta,\mu)$
denotes the statistical expectation value of
the operator $\hat A$ in the reference system $\hat H_0$.

\subsection{Linked cluster theorem at finite temperature}
The terms in Eq.~\eqref{eqn:SExpansion} can be expressed as the exponential
of a subset, consisting only of connected (linked)
terms\,\cite{lancaster_quantum_2014}. Thus,
\begin{equation}
  \Delta\Omega = -\frac1\beta \sum_{n=1}^\infty{(-1)}^n \\
    \int\limits_{
      0<\tau_1<\ldots<\tau_n<\beta
    }{\rm d}\tau_1\ldots{\rm d}\tau_n
    \big\langle \hat H_1(\tau_n)\ldots \hat H_1(\tau_1)\big\rangle'_0\,,
  \label{eqn:FreeEnergyDifference}
\end{equation}
where $\langle\hat A\rangle'_0$ denotes the statistical expectation value
of $\hat A$
in the reference system, restricted to connected Wick contractions only.

\subsection{Finite temperature MP2}
Given the finite temperature Hartree--Fock (HF) operator and the
interacting Hamiltonian in the canonical HF basis
\begin{align}
  \hat H_0&= \sum_p\varepsilon_p\hat c^\dagger_p\hat c_p\,, \\
  \hat H&=
  \sum_{pq} h^p_q\hat c^\dagger_p\hat c_q
  + \frac12\sum_{pqrs}
    V^{pq}_{sr}\hat c^\dagger_p\hat c^\dagger_q\hat c_r\hat c_s\,,
\end{align}
\Eq{eqn:FreeEnergyDifference} can be evaluated, for instance,
up to second order, arriving at finite temperature
M\o ller--Plesset theory for the correlation grand potential 
\begin{equation}
\Delta\Omega^{(2)}
  = -\sum_{abij}\left(
    \frac1{\Delta^{ab}_{ij}}
    +\frac{
      \e^{-\beta\Delta^{ab}_{ij}} - 1
    } {
      \beta(\Delta^{ab}_{ij})^2
    }\right)
    \frac12\,f^{ab}_{ij}(V^{ij}_{ab} - V^{ji}_{ab}) V^{ab}_{ij}
\end{equation}
with the shorthand notations
$\Delta^{ab}_{ij}=\varepsilon_a+\varepsilon_b-\varepsilon_i-\varepsilon_j$,
\begin{equation}
  f_p =\frac{
    \e^{-\beta(\varepsilon_p-\mu)}
  }{
    1+\e^{-\beta(\varepsilon_p-\mu)}
  }, \quad
  f^p = 1-f_p\,,
\end{equation}
and
$f^{ab\ldots}_{ij\ldots}=f^af_if^bf_j\ldots$ Note that
there is, in principle, no distinction between hole and particle indices.
For reference Hamiltonians other than Hartree--Fock, expressions
can be found in Ref.~\citenum{santra_2017}.

\subsection{Thermal expectation values}
\label{ssc:Observables}
Derivatives of the log-partition function with respect to
$\beta$ or $\mu$ yield
central statistical moments of the Hamiltonian $\hat H$ and of
the number operator $\hat N$, respectively
\begin{align}
  \langle\hat H\rangle
    &= \frac{\partial\log\mathcal Z(\beta,\mu)}{-\partial\beta}
  & \langle\hat N\rangle
    &= \frac{\partial\log\mathcal Z(\beta,\mu)}{\beta\partial\mu} \\
  \langle\Delta^2\hat H\rangle
    &= \frac{\partial^2\log\mathcal Z(\beta,\mu)}{(-\partial\beta)^2}
  & \langle\Delta^2\hat N\rangle
    &= \frac{\partial^2\log\mathcal Z(\beta,\mu)}{(\beta\partial\mu)^2}
\end{align}
and so forth, with $\Delta\hat A = \hat A-\langle\hat A\rangle$.

\bibliography{FT-MBPT}

\end{document}